\documentclass[graybox]{svmult}

\usepackage{mathptmx}       
\usepackage{helvet}         
\usepackage{courier}        
\usepackage{type1cm}        
\usepackage{makeidx}         
\usepackage{graphicx}        
\usepackage{multicol}        
\usepackage[bottom]{footmisc}

\makeindex             

\begin{document}

\title*{A new search for variability-selected active galaxies within the VST SUDARE-VOICE survey: the Chandra Deep Field South and the SERVS-SWIRE area
}
\author{Falocco S., De Cicco D., Paolillo M., Covone G., Longo G., Grado A., Limatola L., Vaccari
M., Botticella M.T., Pignata G., Cappellaro E., Trevese D., Vagnetti F., Salvato M., Radovich
M., Hsu L., Brandt W.N., Capaccioli M., Napolitano N., Baruffolo A., Cascone E., Schipani P.}
\institute{S. Falocco, De Cicco D., Paolillo M., Covone G., Longo G., Capaccioli M.  \at Physics Department of University Federico II, Via Cintia CAP 80126 
\and Grado A., Limatola L.,Botticella M.T., Napolitano N., Cascone E.,  P. Schipani \at INAF Osservatorio Di Capodimonte Naples, Italy
\and 
Vaccari
M. \at Astrophysics Group, Physics Department, University of the Western Cape, Cape Town, South Africa
\and Radovich M., Cappellaro E., Baruffolo A. \at INAF- Osservatorio di Padova, Italy
\and Hsu L., Salvato M. \at Max Plank Institute fur Extraterrestrische Physik, Garching, Germany
\and
Pignata G. \at Departamento de Ciencias Fisicas, Universidad Andres Bello, Santiago, Chile
\and 
Trevese D. \at Department of Physics University La Sapienza Roma, Italy
\and Vagnetti F. \at Department of Physics University Tor Vergata Roma, Italy
\and Brandt W.N. \at Department of Astronomy and Astrophysics, The Pennsylvania State University, University Park, PA 16802, USA
}

%
\maketitle

\abstract{This work makes use of the VST observations to select variable sources. We use also the IR photometry, SED fitting and X-ray information where available to confirm the nature of the AGN candidates. The IR data, available over the full survey area, allow to confirm the consistency of the variability selection with the IR color selection method, while the detection of variability may prove useful to detect the presence of an AGN in IR selected starburst galaxies. }

\section{Aims and method}
\label{sec:1}
 The luminosity of virtually all AGN varies at every wavelength (see e.g. \cite{kawaguchi,paolillo2004,garcia-gonzalez2014,ulrich1997} and  references therein),
 thus making variability one of the most distinctive properties of these sources. 
The variability selection method assumes that all AGN vary intrinsically in the observed band, without requiring assumptions on the spectral shape, colours, and/or spectral line ratios.

In the work described in this proceeding we aim at constructing a new variability-selected AGN sample exploiting the data from the ongoing survey performed with the VST (VLT Survey Telescope), see \cite{falocco2015} for details.  
We make use of data in the r band from the SUDARE-VOICE survey performed with the VST telescope (\cite{botticella2013}, Cappellaro et al. in prep.).
In a companion contribution by De Cicco et al. in this volume (also see \cite{decicco2014}, hereafter Paper I) we focused on the COSMOS region. Here we examine two fields around  the CDFS region, that we label CDFS1 and CDFS2.
We examined a total of 27 VST epochs for the CDFS1 and 22 epochs for the CDFS2 spanning five and four months respectively and covering an area of two square degrees. 
The data reduction and the analysis were performed using the procedure explained in  Paper I (and in De Cicco et al., this volume) consisting in identifying as candidate AGNs all sources whose lightcurve showed an excess variability of 3 r.m.s. from the average variability of all sources with similar magnitude.  
 To validate our catalogue of
variable objects we exploited SWIRE by \cite{lonsdale2004} and
SERVS by \cite{mauduit2012}. We also used SED (Spectral Energy Distribution) classification given in \cite{hsu2014} and \cite{rowan-robinson2013}.

\section{Results and discussion}
\label{sec:3}

We obtained a sample of 175 sources that we investigated further in detail analysing the diagnostics described below. 
12\% of the selected sample are classified as SN, based on both visual inspection of the light-curves and template fitting by the SUDARE-I collaboration (Botticella et al., this volume and Cappellaro et al. in prep.).

We used the information contained in \cite{hsu2014}, to extract X-ray and SED data for our variable candidates located within the ECDFS area.
  There are only 15 sources in common between the sample presented in \cite{hsu2014} and our selected sample.
 The 15
common sources belong to the CDFS1 which encloses the ECDFS. 
  Twelve of the 15 common sources are detected in the X-rays and their SEDs require a strong AGN contribution (in particular in the NIR part of the spectrum).
 All these sources have also been identified as non-SN on the basis of the inspection of their lightcurves. 
The remaining three sources are non-detected in the X-rays and their best-fit SED template shows no evidence for a significant AGN contribution. These three sources were identified as SN according to their lightcurves. Therefore, we conclude that they are SN explosions in normal galaxies.

\begin{figure*}[t]
\includegraphics[width=8.5cm]{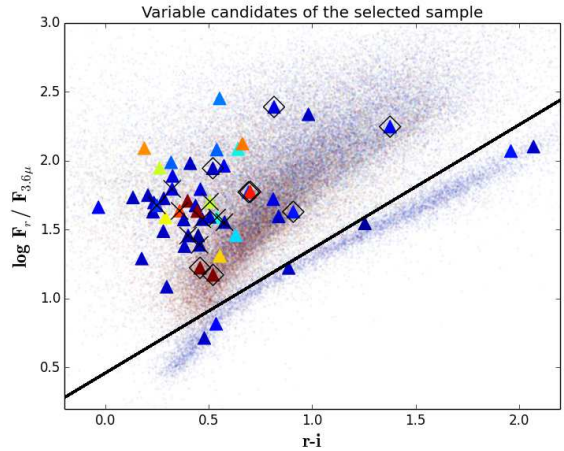}
\caption{\small{Flux (F$_{\lambda}$) ratios between the $r$ band and 3.6 $\mu$m versus $r-i$ colour. 
 Small points: SERVS+SWIRE 82254 sources. Triangles: 57 sources in common with the selected sample. Diamonds: SN. Crosses: X-ray detected sources. The colours indicate the increasing stellarity, from red (extended sources) to blue (pointlike). The solid line separates the stellar sequence and the non-stellar region.}}
\label{36ri}       
\end{figure*}

We validate our catalogue of variable objects with the overlapping surveys
SWIRE \cite{lonsdale2004} and SERVS \cite{mauduit2012} which provide data in the 3.6, 4.5, 5.6, 8, 24, 70, 160 $\mu m$ bands, and in the U, g, r, i, z
filters.
 In Fig. \ref{36ri} we compare the $r-i$ versus the 3.6 $\mu m$ to $r$ band flux ratio of our variable candidates with the SERVS+SWIRE source catalog. This diagram has been proposed by \cite{rowan-robinson2013} to separate stars from galaxies. 
The populations represented in the plot are segregated into two regions: stars and extragalactic objects. Figure \ref{36ri} shows that six of 57 variable objects are along the stellar sequence, likely variable stars.

\begin{figure*}
\includegraphics[width=8.5cm]{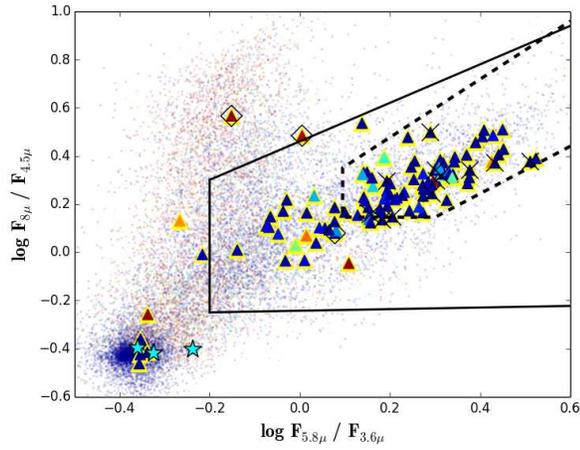}
\caption{\small{Flux (F$_{\lambda}$) ratio (logarithmic) at 5.8 and 3.6 $\mu m$ versus flux ratio at 8 and 4.5 $\mu m$. Small points: SERVS+SWIRE 18436 sources; Triangles (enclosed in yellow edges): 115 sources in common with the selected sample. Cyan stars: stars; Diamonds: SN. Crosses: X-ray detected sources. Colour code as in Fig. \ref{36ri}. The solid line is the Lacy region and the dashed line the Donley region (see text).}}
\label{lacy}       
\end{figure*}

We further make use of the mid-IR colors in order to confirm the identification of our AGN candidates.
Figure \ref{lacy} shows the diagnostic developed by \cite{lacy2004}.
Due to the different dust content and temperature, normal galaxies, starforming galaxies and AGNs occupy different regions of this diagram. This allows us, as shown 
in \cite{lacy2004}, to define an empirical wedge (solid line in Fig. \ref{lacy}) which encloses a large fraction of the AGN population.
Out of the 115 sources of the selected sample represented in the plot, 103 lie within the Lacy wedge, supporting their AGN nature. 
We also note from Fig. \ref{lacy} that the average stellarity
  index of the variable candidates inside the Lacy region decreases
  towards the left side of the diagram, where the contamination is more severe.
According to \cite{trevese2008}, many of such sources are low ionisation narrow emission regions (LINERs).
To improve the purity \footnote{We define the purity as the number of confirmed AGN divided by the number of AGN candidates} of the IR selected AGN sample and to reduce the starburst contamination to IR-selected AGN samples, \cite{donley2012} defined a more restrictive criterion, which is shown in Figure \ref{lacy} as a dashed line.  
The majority of pointlike sources lie within the Donley wedge, strengthening the view that the Donley region is occupied prevalently by AGN-dominated galaxies. 

\section{Conclusions}
\label{sec:4}
We identified 175 candidates selected through variability using VST observations in the CDFS. 
To validate the sample, we used information available both within the VST-SUDARE consortium and in the literature.
The total number of candidates for which we could employ the diagnostics discussed in the previous section is 137 out of a total 175 candidates in the selected sample.
We found 103 confirmed AGN (by at least one diagnostic of those explored in the previous section), that is 75\% of the 137 candidates with ancillary data and 59\% of the selected sample. 
As expected, contaminants are mainly stars and SN: the stars constitutes 3 \% of the selected sample of 175 candidates, while the SN constitute 12 \%.

In conclusion the purity of our sample of optically variable sources is 75\%, close to the 80\% obtained for the COSMOS field in Paper I.
The completeness of the variability-selected survey presented in this work is 22 \% (computed with respect to the IR selection of \cite{donley2012}).
In Paper I, the completeness (computed with respect to X-ray samples) has been estimated to be 15 \% for a 5 months baseline; the two results are thus in broad agreement considering that they are estimated with respect to different reference populations. 
These completeness levels can improve extending the monitoring baseline, as pointed out in previous papers (e.g. \cite{sesar2007} and Paper I).
The new observations which are currently being acquired for the COSMOS field (P.I: G. Pignata) with VST will allow us to directly compare these results using a 3-year long monitoring baseline.

%



%
%
%
\biblstarthook{
}

\end{document}